\documentclass[fleqn,10pt]{wlscirep}
\usepackage[utf8]{inputenc}
\usepackage[T1]{fontenc}
\usepackage{ulem}

\usepackage{upgreek}
\usepackage{hyperref}% add hypertext capabilities

\title{Proton beam quality enhancement by spectral phase control of a PW-class laser system}

\author[1,2,*]{T.~Ziegler}
\author[1]{D.~Albach}
\author[1,2]{C.~Bernert}
\author[1]{S.~Bock}
\author[1,2]{F.-E.~Brack}
\author[1,2]{T.~E.~Cowan}
\author[3]{N.~P.~Dover}
\author[1,2]{M.~Garten}
\author[1,2]{L.~Gaus}
\author[1]{R.~Gebhardt}
\author[1,2]{I.~Goethel}
\author[1]{U.~Helbig}
\author[1]{A.~Irman}
\author[3]{H.~Kiriyama}
\author[1]{T.~Kluge}
\author[3]{A.~Kon}
\author[1]{S.~Kraft}
\author[1]{F.~Kroll}
\author[1]{M.~Loeser}
\author[1]{J.~Metzkes-Ng}
\author[3]{M.~Nishiuchi}
\author[1,+]{L.~Obst-Huebl}
\author[1]{T.~Püschel}
\author[1,2]{M.~Rehwald}
\author{H.-P.~Schlenvoigt}
\author[1,2]{U.~Schramm}
\author[1]{K.~Zeil}

\affil[1]{Helmholtz-Zentrum Dresden - Rossendorf, Institute of Radiation Physics,~01328 Dresden, Germany}
\affil[2]{Technische Universität Dresden, 01069 Dresden, Germany}
\affil[3]{Kansai Photon Science Institute, National Institutes for Quantum and Radiological Science and Technology, Kyoto 619-0215, Japan}
\affil[+]{Current address: Lawrence Berkeley National Laboratory, California 94720, USA}

\affil[*]{t.ziegler@hzdr.de}

\begin{abstract}
We report on experimental investigations of proton acceleration from solid foils irradiated with PW-class laser-pulses, where highest proton cut-off energies were achieved for temporal pulse parameters that varied significantly from those of an ideally Fourier transform limited (FTL) pulse.
Controlled spectral phase modulation of the driver laser by means of an acousto-optic programmable dispersive filter enabled us to manipulate the temporal shape of the last picoseconds around the main pulse and to study the effect on proton acceleration from thin foil targets.
The results show that applying positive third order dispersion values to short pulses is favourable for proton acceleration and can lead to maximum energies of 70\,MeV in target normal direction at 18\,J laser energy for thin plastic foils, significantly enhancing the maximum energy compared to ideally compressed FTL pulses.
The paper further proves the robustness and applicability of this enhancement effect for the use of different target materials and thicknesses as well as laser energy and temporal intensity contrast settings.
We demonstrate that application  relevant proton  beam  quality was reliably achieved over many months of operation with appropriate control of spectral phase and temporal contrast conditions using a state-of-the-art high-repetition rate PW laser system. 
\end{abstract}

\begin{document}

\flushbottom
\maketitle
\thispagestyle{empty}

%\noindent Please note: Abbreviations should be introduced at the first mention in the main text – no abbreviations lists. Suggested structure of main text (not enforced) is provided below.
%----------------------------------------------
\section*{Introduction}
%----------------------------------------------
Laser-driven ion acceleration \cite{Daido2012, Macchi2013} as a very compact accelerator technology with remarkable beam properties has been associated with a multitude of medical \cite{Malka2004, Masood2017}, scientific \cite{Patel2003, Romagnani2005, Obst-Huebl2018a,NJP_Roadmap2020} and technical \cite{Roth2001, Barberio2017, Barberio2018} applications for several years now. Realizing those applications turned out to be highly complex requiring a sophisticated level of control on the laser plasma interaction process, which determines the beam quality and energy.
Key to any progress on that matter is a detailed understanding of the underlying physics as well as appropriate technical control and metrology of the acceleration process, which have therefore been extensively studied both experimentally and theoretically over the last 20 years.\\
% TNSA description %
Target normal sheath acceleration (TNSA) is the most robust and widely understood acceleration regime, and has therefore received particular attention in the context of applications.
It describes the generation of electric space-charge fields ($\gtrsim$\,TV/m), driven by laser-accelerated prompt front-side electrons, by which particles from a contaminant layer at the target rear side get ionized and accelerated to energies of several tens of MeV per nucleon.
Employing dedicated laser-target configurations (e.g. ultra-thin, low density, special shape targets) allowed for control and establishment of optimized TNSA-based as well as other advanced acceleration regimes whereby recent experiments have demonstrated that combinations of those or hybrid schemes show huge potential \cite{Higginson2018, Hilz2018, Ma2019}.
These efforts are complemented by a variety of laser pulse parameter scans (e.g. energy, duration, shape, temporal contrast of the pulse) to determine the optimal laser proton accelerator performance \cite{Fuchs2006, Kaluza2004, Zeil2012_nature-com, Brenner2014, Obst2018, Tayyab2018}.\\
Yet, highest proton energies were mainly achieved with high intensity long-pulse lasers delivering only a few shots per day which prevents application-relevant high average currents\cite{Wagner2016,Higginson2018}.
Ultra-short pulse laser systems (few tens of femtoseconds pulse duration) with high repetition rate (up to 10\,Hz) hold the promise to bridge this gap\cite{Kim2016} and given the recent progress in laser technology, numerous facilities worldwide\cite{Papadopoulos2016,Schramm2017a,Kiriyama2018,Sung2017,Gan2017,Nakamura2017} approach or even surpass the PW-level with on target intensities between $10^{21}$ and $10^{22}$ W/cm$^2$.
%Ultra-short pulse laser systems (few tens of femtoseconds pulse duration) with high repetition rate (up to 10\,Hz) could bridge this gap and given the recent progress in laser technology, numerous facilities worldwide\cite{Papadopoulos2016,Schramm2017a,Kiriyama2018,Sung2017,Gan2017,Nakamura2017} approach or even surpass the PW-level with on target intensities between $10^{21}$ and $10^{22}$ W/cm$^2$.
Furthermore, these sources provide additional options for control, modifications and diagnostics being of particular importance for the characterization of laser pulse parameters in focus at these intensities.
Upon main pulse arrival, the real plasma conditions due to pre-pulses or spatio-temporal couplings may differ significantly from those assumed in idealized theoretical models.
In view of exploiting the full potential of laser driven ion accelerators, on-shot diagnostics and feedback routines based on advanced computing methods, like already applied for wakefield accelerators\cite{shalloo2020}, might also become an option.\\
We experimentally demonstrate that actively manipulating the temporal pulse shape of the driver laser significantly enhances the proton acceleration performance using a state-of-the-art PW ultra-short pulse system.
In a series of experiments under well-controlled contrast conditions with different target materials and thicknesses as well as laser energy and temporal intensity contrast configurations, we found that proton cut-off energies and particle numbers were consistently enhanced by changing the temporal laser profile from a Fourier transform limited (FTL) to an asymmetric pulse shape.
%With optimized settings we were able to routinely deliver maximum proton energies around 60\,MeV which corresponds to $\sim$\,3.4\,MeV per Joule laser energy on target.
With optimized settings we were able to routinely deliver maximum proton energies around 60\,MeV.
%Compared to the nominal settings, thus an significant enhancement of the maximum proton energies and particle numbers can be achieved.
Compared to the nominal settings, thus an average enhancement of the maximum proton energies of $\sim 37\,\%$ could be achieved.
Based on the simplicity of the method and the long-term stability of our results, we believe that this optimization method is universally applicable to other laser systems with particular importance when operating in the PW regime.\\
%%%
\section*{Results}

\begin{figure*}[htb]
\includegraphics[width=0.95\textwidth]{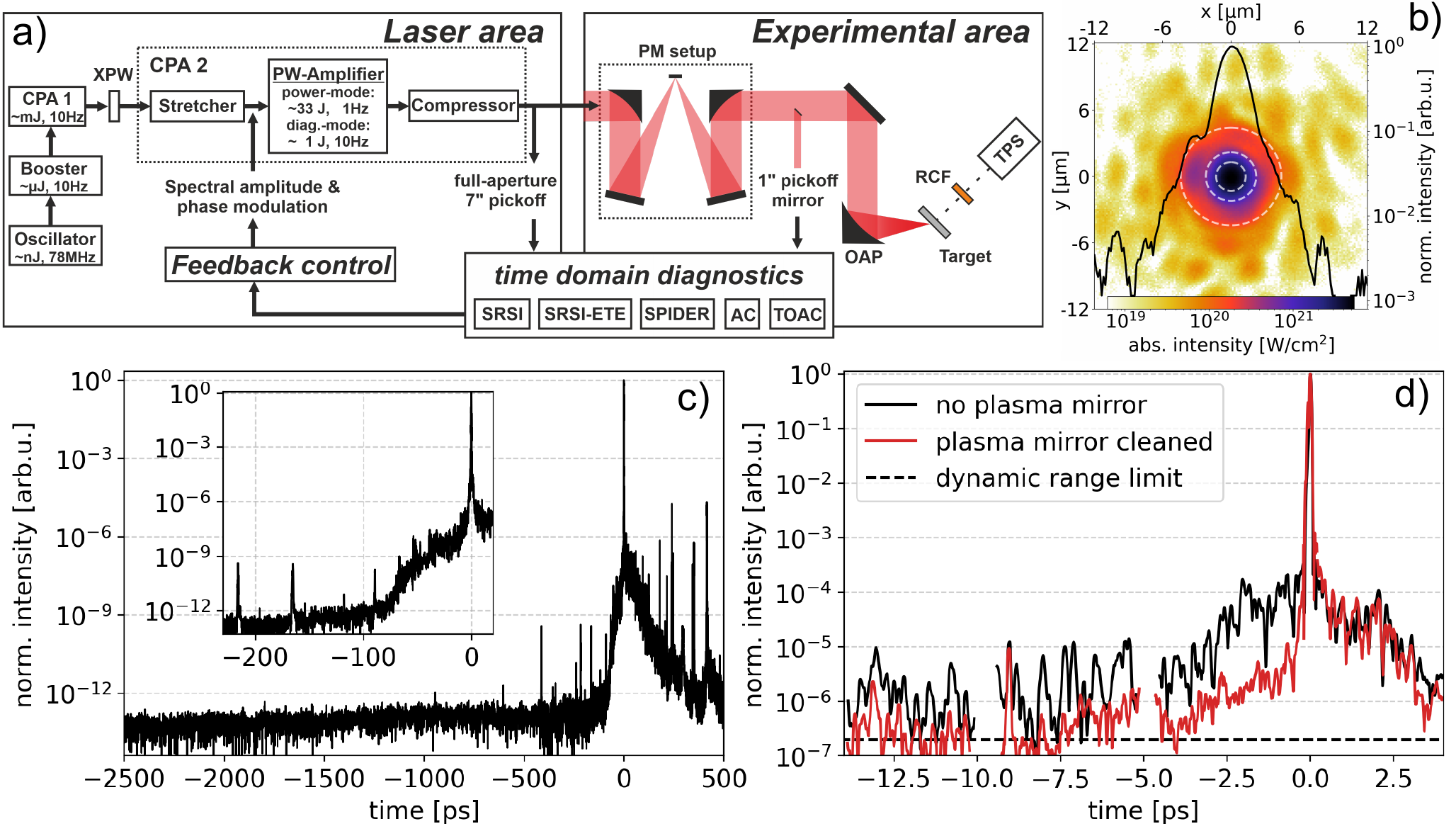}
\caption{\label{fig:setup}
a) Illustration of the DRACO PW laser, the experimental area, the two pick-off ports and the different diagnostics for time domain measurements of the laser pulse.
b) Magnified DRACO focal spot measurement at the experimental area with logarithmic color scale for absolute intensities. The black line represents a normalized horizontal line out of the focal intensity distribution, the white dashed circles represent the FWHM, 2$\sigma$ and 4$\sigma$ area.
c) + d) Temporal intensity contrast of the DRACO laser on the: c) ns-range (inset: 100\,ps), measured with scanning TOAC (\textit{SequoiaHD}), d) ps-range for intrinsic (black) and PM cleaned (red) contrast conditions, measured with single-shot time extended self-referenced spectral interferometry technique \cite{Oksenhendler2017} (SRSI-ETE).}
\end{figure*}

The presented experiments were carried out at the ultra-short pulse laser DRACO \cite{Schramm2017a} at the Helmholtz-Zentrum Dresden\,-\,Rossendorf (HZDR). DRACO is a dual beam double CPA (chirped pulse amplification) Ti:Sa laser system, designed to deliver 30\,J within 30\,fs on target with 1\,Hz repetition rate. A simplified sketch of the laser system alongside the experimental setup can be found in FIG. \ref{fig:setup}a).

%Pulses from a passively Kerr-lens mode-locked oscillator get stretched, amplified to the mJ-level and re-compressed in the first CPA stage before they enter a second CPA stage consisting of an all-reflective Öffner-stretcher, regenerative and several multi-pass amplifiers and the vacuum grating-compressor.

The temporal pulse structure of DRACO was characterized with rigorous care and a broad variety of scanning and single-shot diagnostics.
This includes second and third order autocorrelators (AC and TOAC), field auto-correlation methods like self-referenced spectral interferometry (SRSI \& SRSI-ETE) and spectral phase interferometry for direct electric-field reconstruction (SPIDER) at different positions (vacuum compressor output \& just before final focusing) and pick-off methods (full-aperture\,$\simeq$\,7" \& 1"~mirror) within the laser chain and for different energy settings (diagnostic-mode \& power-mode which corresponds to non-pumped (1\,J) or fully-pumped main-amplifiers (33\,J), respectively).
Temporal pulse contrast optimization is achieved by a series of fast pockels cells with optimized timing structure and minimal timing jitter and XPW filtering between the two CPA stages yielding an intensity contrast ratio better than 10$^{-12}$ up to -100\,ps prior to the main pulse as depicted in FIG.~\ref{fig:setup}c).
The inset shows the rise of the coherent pedestal at -75\,ps which persists at 10$^{-8}$ until -10\,ps.
The few visible pre-pulse-like signatures between -500\,ps and -100\,ps can partially be identified as measurement artefacts typical for TOAC, reflecting the existence of post-pulses generated by internal reflections in remaining planar transmission optics (e.g. amplifier crystals). Dominantly the signatures represent the conversion of such post-pulses into pre-pulses by non-linear processes associated with the accumulated B-integral in the amplifier chain \cite{Didenko2008, Kiriyama2020}.
Remaining below a level of 10$^{-9}$ they can be further suppressed on-demand by inserting a re-collimating single plasma mirror (PM) setup installed close to the target.
%Temporal pulse contrast optimization is achieved by fast pockels cells with optimized timing structure and minimal timing jitter and XPW filtering between the two CPA stages yielding an intensity contrast ratio better than 10$^{-12}$ up to -100\,ps prior to the main pulse as depicted in FIG. \ref{fig:setup} c).
%The inset shows the rise of the coherent pedestal at -75\,ps which persists at 10$^{-8}$ until -10\,ps.
%All visible pre-pulses between -500\,ps and -100\,ps could be identified to originate from post-pulses generated by internal reflections at transmitting optics (crystals, pockels cells, compressor windows) which couple into real pre-pulses with a slight shift in time via non-linear processes associated with the B-integral accumulated in the laser chain \cite{Didenko2008, Kiriyama2020}.
%They can be suppressed on-demand by introducing a re-collimating single plasma mirror (PM) setup installed inside the experimental chamber.
The PM yields an enhancement of the intrinsic temporal contrast by almost two orders of magnitude resulting in an intensity ratio better than 10$^{-5}$ at -1\,ps prior to the main pulse as depicted in FIG.~\ref{fig:setup}d) for the ps time window.
Sub-ps pulse optimization is achieved by controlling the spectral amplitude and phase of the coherent portions of the laser beam.
Therefore, two acousto-optic programmable dispersive filters (AOPDFs), namely \textit{Mazzler}\cite{Oksenhendler2006} and \textit{Dazzler}\cite{Verluise2000} from Fastlite/AmplitudeTechnologies, are incorporated in each CPA stage to maintain the desired spectral shape and, respectively, the spectral phase components by pre-compensation of higher order residual phase terms acquired by the laser pulse while propagating through the laser chain.
\newline
After the PM, the wave-front corrected laser pulse with a total remaining energy of 18\,J is focused by an \mbox{F/2.3} parabola to a full width at half maximum (FWHM) spotsize of 2.6\,$\mathrm{\upmu m}$ yielding peak intensities of $5.4 \times 10^{21}\,\mathrm{W/cm}^{2}$.
The high spatial quality of the focused laser beam can be seen in FIG.~\ref{fig:setup}b), where the dashed circles represent the FWHM, $2\mathrm{\upsigma}$ and $4\mathrm{\upsigma}$ area containing 35\,\%, 58\,\% and 82\,\% of the total laser energy, respectively.
The laser pulses irradiated the targets at an incidence angle of 45$^\circ$ with p-polarization.\\
The main particle diagnostic to detect and analyze the accelerated ion beam was a multi-channel plate equipped Thomson parabola spectrometer (TPS) aligned to the target normal direction providing an energy dependent resolution of 5\,\% with a minimum detectable proton energy of 7\,MeV.
For some selected shots stacks of calibrated radiochromic films (RCF) were inserted at a distance of 55\,mm behind the target allowing for proton beam profile characterization, absolute particle number calibration and complementary maximum energy detection.\\
\begin{figure}[htb]
\centering
\includegraphics[width=0.7\textwidth]{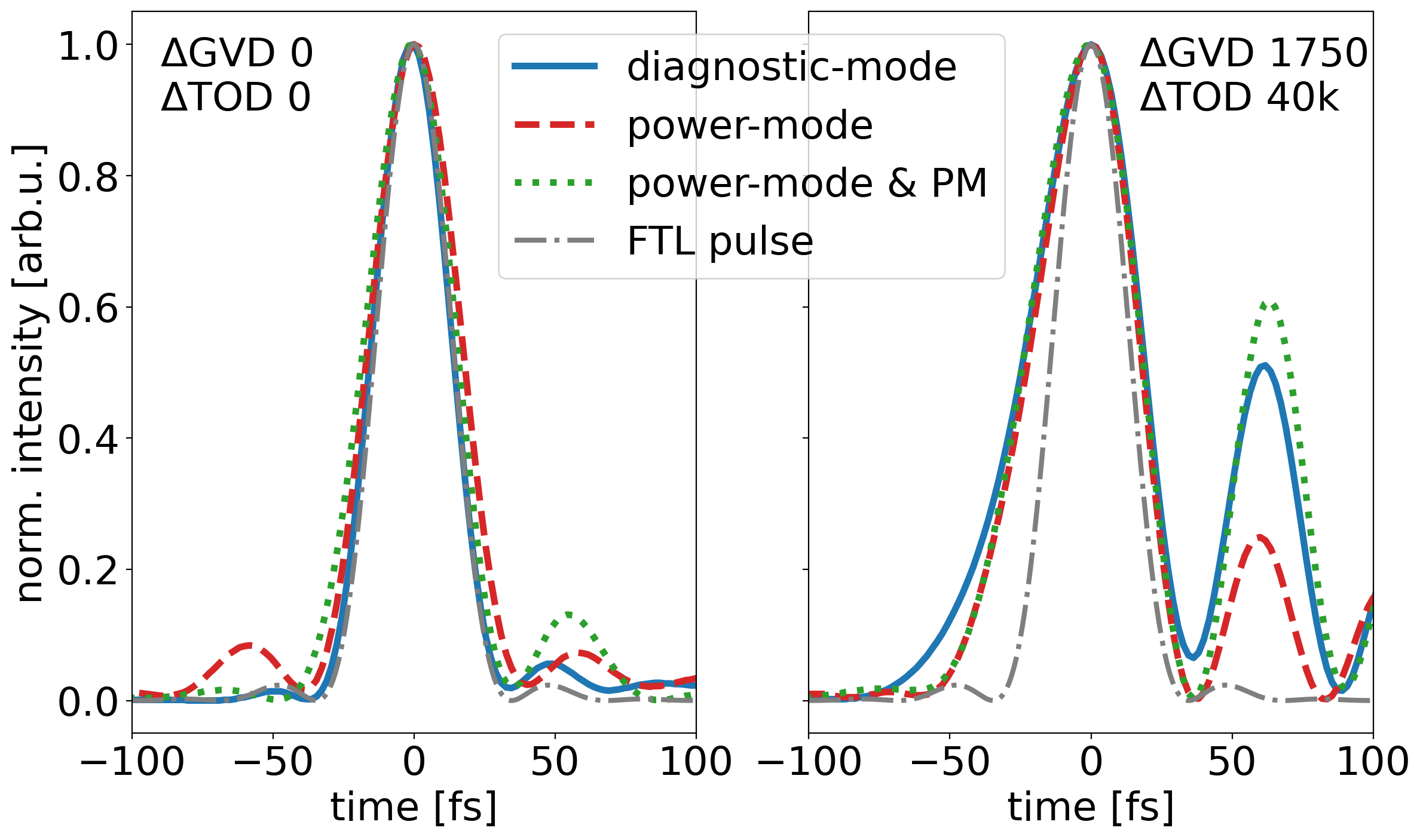}
\caption{\label{fig:draco-spider-all}  Temporal laser pulse shapes retrieved from SPIDER measurements for different spectral phase configurations (left: after automatic \textit{Dazzler} Feedback-loop, right: manual phase manipulation $\Delta$GVD 1750\,fs$^2$, $\Delta$TOD 40k\,fs$^3$). While different laser energy configurations (diagnostic mode - solid blue, power mode - dashed red, power mode \& PM - dotted green) show consistent pulse shapes, the two spectral phase settings significantly differ between an almost ideally compressed FTL pulse (dash-dotted grey) and an asymmetrical, slightly longer pulse with shallow rising edge and shifted pre- and postpuls distribution.
%Further potential influence of these phase manipulations on pulse contrast is beyond the resolution of the measurement.
}
\end{figure}
%\section{Results}
%The experimental measurements were performed for different target materials and thicknesses while adjusting the Dazzler-settings and thereby changing the temporal pulse shape of the laser.
\newline
For the experimental measurements we manually varied the spectral phase terms group velocity dispersion (GVD) and third order dispersion (TOD) by presetting the according values in the \textit{Dazzler} device, enabling us to individually adjust the instantaneous frequencies of the electric field and thus the temporal shape of the laser pulse.
First, we ensured that the automatic \textit{Dazzler} feedback loop produces a flat phase over the entire laser spectrum providing almost ideal FTL pulses for all the different laser energy and PM configurations, examples of which are shown by the SPIDER measurements on the left in FIG.~\ref{fig:draco-spider-all}.
Simultaneous measurements performed with the different redundant time domain diagnostics and pick-off ports delivered consistent results, thus all relative phase changes introduced in the following can be referenced to a 30\,fs FWHM near Gaussian pulse shape (standard case).
%Make sure that manipulation is valid for all laser configurations and is not perturbed by ... or that we are not compensating different phase terms introduced by optical components when they experience the full amplification fluence.\\
On that basis, a pure GVD change preserves the symmetric shape but stretches the pulse in time resulting in a reduction in peak intensity.
A pure modification of the TOD leads to an asymmetric pulse shape, identified by a shallow rising and sharp falling edge (or vice versa depending on sign) and reduction in peak intensity due to frequency components being shifted away from the main pulse which results in post- or pre-pulse generation and reduction.
Measurements with the different time domain diagnostics confirm the described effects of spectral phase changes on the temporal pulse shape (c.f. FIG.~\ref{fig:draco-spider-all}).
%\newline
%Having confirmed the effects of spectral phase changes on the temporal intensity distribution close to the peak, we systematically investigated the influence of those changes on proton acceleration for 400\,nm Formvar targets.

\begin{figure}[htb]
\centering
\includegraphics[width=0.7\textwidth]{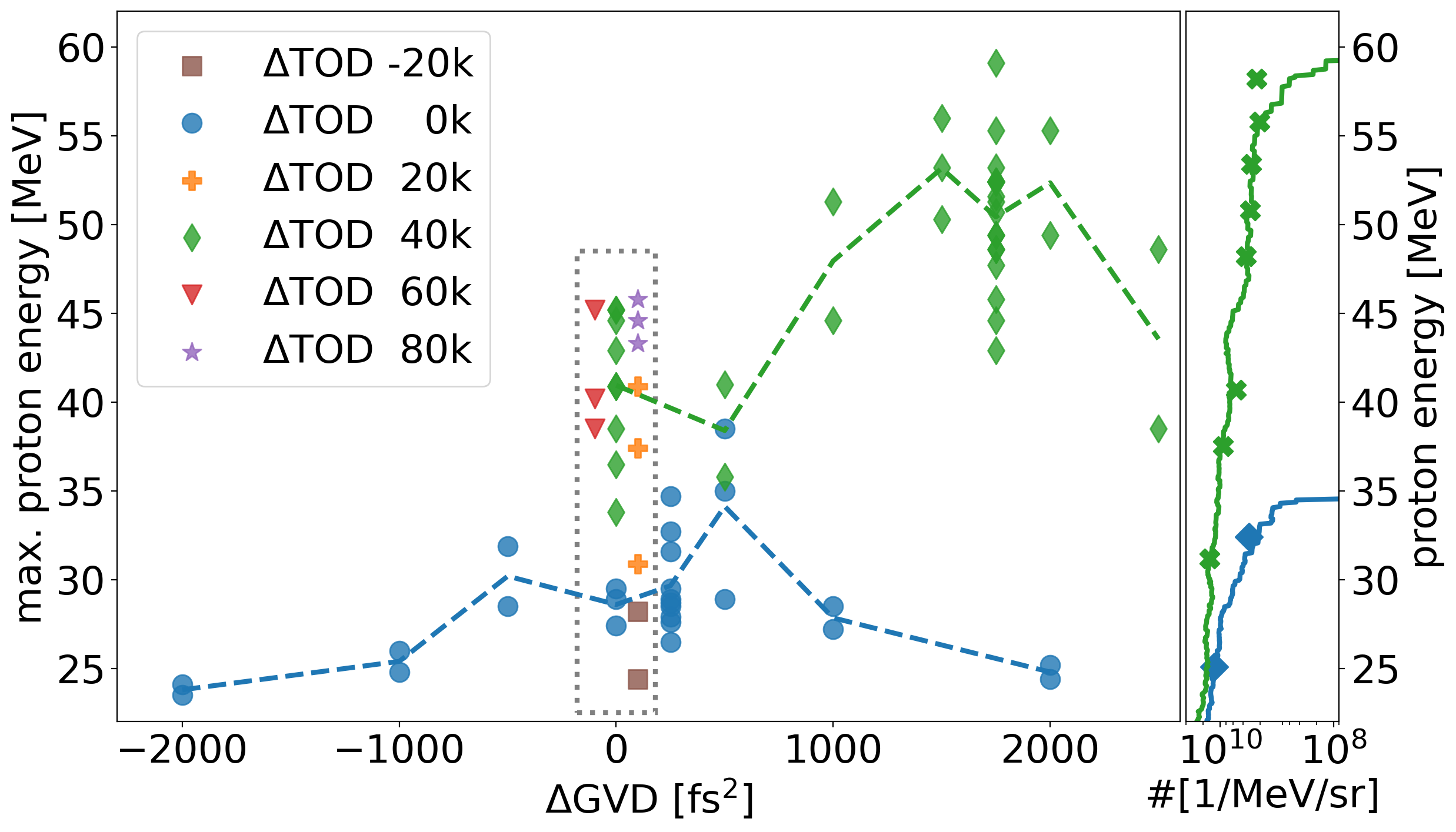}
\caption{\label{fig:gvd_rcf} Maximum proton energies from 400\,nm Formvar targets for different GVD and TOD Dazzler values and PM cleaned contrast. Each marker represents one single shot, dashed lines connect mean values. While the maximum energy for the standard settings ($\Delta$GVD=0\,fs$^2$, $\Delta$TOD=0\,fs$^3$) is below 30\,MeV, the optimized conditions (1750\,fs$^2$, 40k\,fs$^3$) yield 60\,MeV and thus an effective doubling of the maximum energy in this case.
The right side plot shows particle numbers from shots with TPS (solid lines) and shots with RCF (individual markers) measurements for the standard (blue) and optimized (green) settings with higher particle numbers in the optimized case.}
\end{figure}

We then systematically investigated the influence of those spectral phase changes on proton acceleration for 400\,nm Formvar targets.
FIG.~\ref{fig:gvd_rcf} shows the resulting cut-off energies and particle numbers for different Dazzler phase term modifications $\Delta$GVD and $\Delta$TOD.
%\textcolor{red}{The values were set directly to the device and can therefore differ from the phase term representation at the final laser focus.}
%
%How the temporal pulse shape variation is effecting the proton acceleration performance with respect to cut-off energies and particle numbers for 400\,nm Formvar targets is shown in FIG.\ref{fig:gvd_rcf}.
While initially keeping the GVD unchanged ($\Delta$GVD\,=\,0\,fs$^2$), we varied the TOD from $-20\mathrm{k}\,\mathrm{fs}^3$ to $+80\mathrm{k}\,\mathrm{fs}^3$ in $20\mathrm{k}\,\mathrm{fs}^3$ steps (represented by different colors inside the dotted rectangle in FIG.~\ref{fig:gvd_rcf}).
%
%While initially keeping the GVD unchanged from the optimal settings (0\,fs$^2$ relative GVD), the TOD was increased from 0\,fs$^3$ to 60$\times$1e3\,fs$^3$ in 20$\times$1e3\,fs$^3$ steps.
Negative TOD values degrade the acceleration performance, whereas positive TOD values generally result in higher proton cut-off energies, which increase from below 30\,MeV to more than 40\,MeV.
However, a clear optimum is not apparent from this data set, especially since we could not further increase the TOD without producing deep and sharp modulations of the laser spectrum, critical for system safety.
%, and shot to shot fluctuation needs to be considered as well.
\newline
%To clarify whether the observed proton energy enhancement can be attributed to the asymmetric shape or the altered length of the laser pulse, we performed an additional GVD scan for TOD values 0\,fs$^3$ and 40k\,fs$^3$.
%For TOD 0\,fs$^3$ the GVD was varied between -2000\,fs$^2$ and +2000\,fs$^2$ without having a strong influence on the maximum proton energies at all.
To clarify whether the observed proton energy enhancement can be attributed to TOD-induced pulse shape modifications or to the simultaneously altered length of the laser pulse, we performed an additional GVD scan for TOD values 0\,fs$^3$ and 40k\,fs$^3$.
For TOD 0\,fs$^3$ the GVD was varied between -2000\,fs$^2$ and +2000\,fs$^2$ without having a comparable large effect on the maximum proton energies.
At $\pm2000\,\mathrm{fs}^2$ cut-off energies drop below 25\,MeV as a result of the reduced laser intensity due to the larger pulse duration. %($\sim10\times$FTL)
Keeping the TOD value fixed at 40k\,fs$^3$, we scanned the GVD between 0\,fs$^2$ and 2500\,fs$^2$ which led to a further energy enhancement for higher GVD values, clearly peaking at 1750\,fs$^2$ with 60\,MeV, followed by a decrease for even higher GVD values.
RCF measurements confirm the TPS results and prove a clear enhancement effect for the optimized spectral phase parameters in terms of particle numbers as well (right side plot in FIG.~\ref{fig:gvd_rcf}).
This results in a laser-to-proton energy conversion efficiency of $\sim 4\,\%$ for protons with kinetic energies of more than 20\,MeV.
The SPIDER measurements on the right in FIG. \ref{fig:draco-spider-all} reveal that the laser pulse in the optimized acceleration case ($\Delta\mathrm{GVD}\,=\,+1750\,\mathrm{fs^2}$, $\Delta\mathrm{TOD}\,=\,+40\mathrm{k}\,\mathrm{fs^3}$) still has a well compressed but asymmetric shape, represented by a shallow rising edge followed tens of fs later by a non-negligible post-pulse structure. % on the linear intensity scale. 
Higher or lower GVD values increase the pulse duration and yield lower cut-off energies as a result of the reduced peak intensity.
% Additionally, changing the TOD values also results in an asymmetric modification of the pre- and post-pulse structure.
%\newline

\begin{figure}[htb]
\centering
\includegraphics[width=0.6\textwidth]{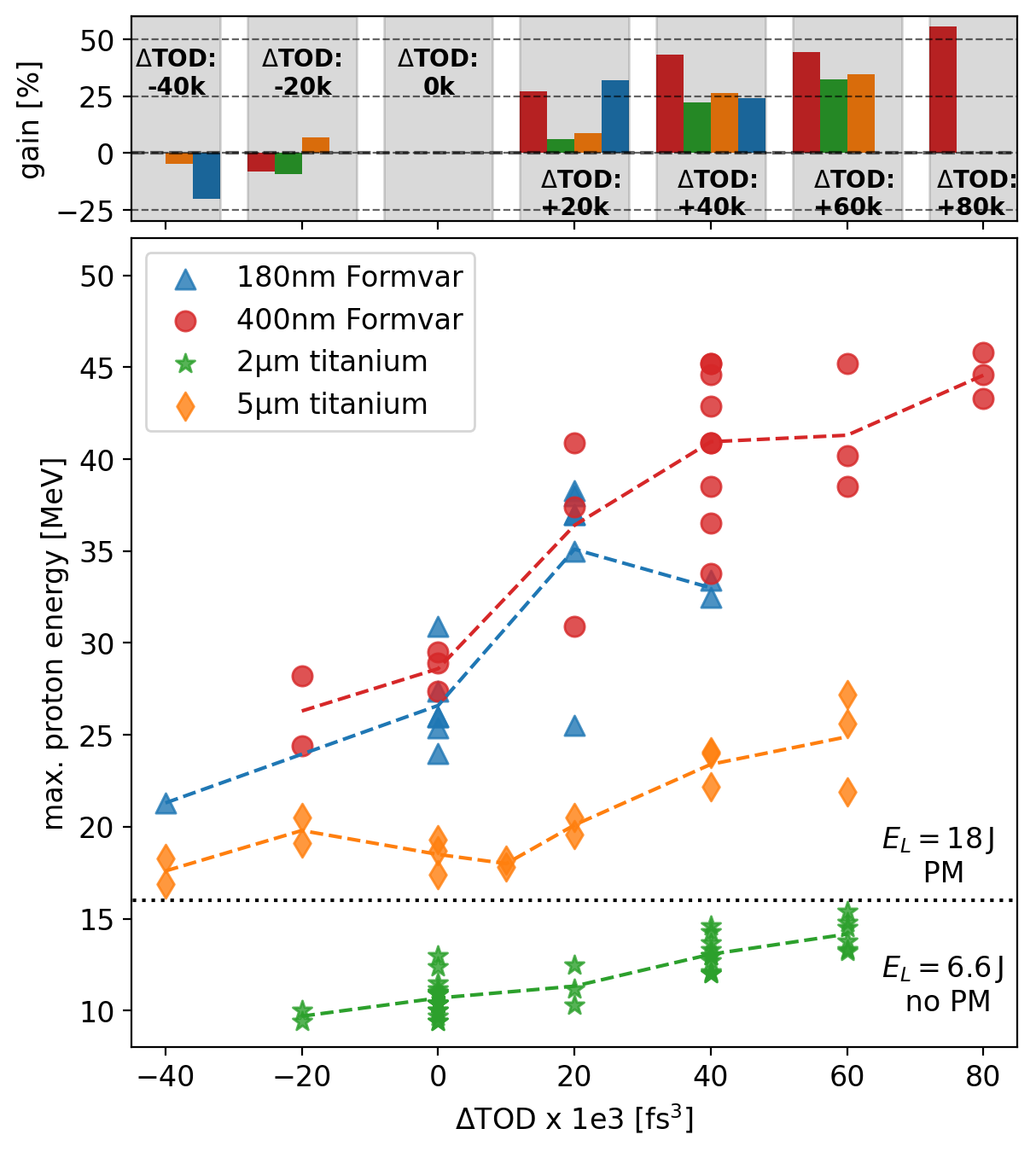}
\caption{\label{fig:tod-scan} Maximum proton energy with respect to $\Delta$TOD ($\Delta$GVD\,=\,0) for different target materials and thicknesses (represented by different markers and colors) as well as on target laser energy ($E_L$) and temporal contrast (PM and no PM) settings.
The upper plot shows the relative energy gain with respect to the standard settings for the different target types and $\Delta$TOD values.}
\end{figure}

\begin{figure}[htb]
\centering
\includegraphics[width=0.95\textwidth]{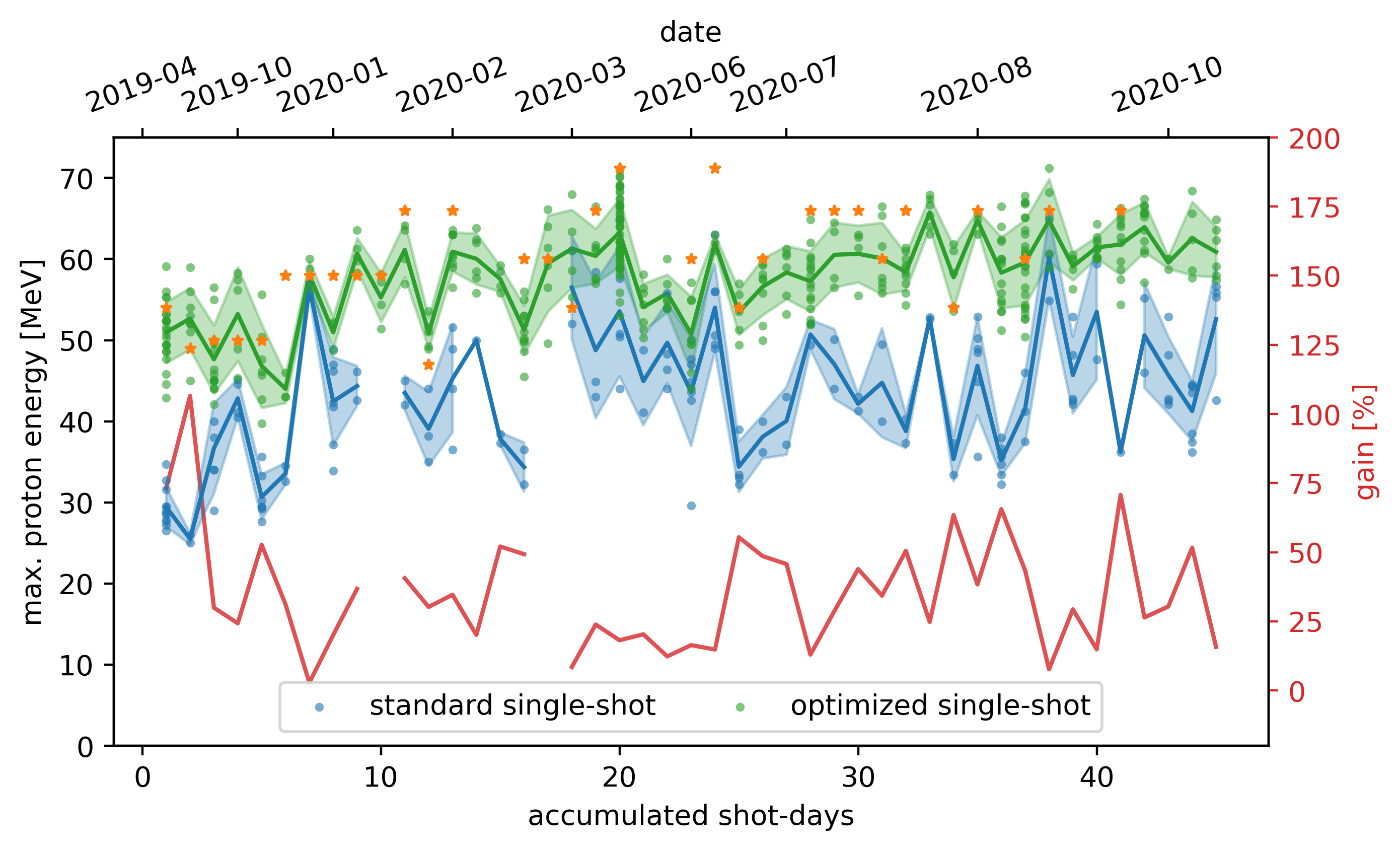}
\caption{
\label{performance}
Long-term stability of the pulse shape optimization induced enhancement effect over a period of more than one year of operation for a total of 575 shots on 45 days of laser proton acceleration experiments. The laser target configuration was the same or very similar to the one described in the text (18\,J pulse energy, plasma mirror cleaned contrast, oblique laser incidence, 200-400\,nm thick Formvar targets).
Compared are the performances obtained for standard (blue dots) and optimized spectral phase settings (green dots) represented by measured proton cut-off energies with TPS and complemented by RCF stack data (orange stars) whenever available. Each marker represents one single shot. The shaded area represents the standard deviation, the solid lines connect the mean of individual data sets serving as guide for the eye. The red solid curve indicates the performance gain in mean energy that was achieved by finding the optimized spectral phase settings.
}
\end{figure}

As the observed gain in energy and particle number is correlated to the change of the TOD values applied, we further studied the stability of this enhancement effect for various other laser-target configurations.
%the asymmetry of the laser pulse (for non-ideal FTL, caused by the applied spectral phase changes)
FIG.~\ref{fig:tod-scan} shows the effect of scanning the TOD while keeping the GVD unchanged (GVD\,=\,0\,fs$^2$) on the maximum proton energy for 180\,nm and 400\,nm Formvar as well as $5\,\mathrm{\upmu m}$ and $2\,\mathrm{\upmu m}$ titanium targets, where in the latter case the PM was removed and the laser energy was reduced to 6.6\,J.
The obtained results reveal that the general trend of the enhancement effect exists for all studied configurations which cover a broad parameter range and hence different initial interaction conditions.
%Although the relative enhancement of the maximum proton energies varies for these different cases, the data show that a $\sim$ 20\% gain is always achievable and an asymmetric pulse shape, induced by positive TOD values, always leads to higher maximum proton energies while lower TOD values decrease the acceleration performance.
Although the relative enhancement of the maximum proton energies varies for these different cases, the data show that a $\sim$ 20\,\% gain is always achievable.
Positive TOD values thereby always lead to higher maximum proton energies while lower TOD values decrease the acceleration performance.
%An appropriate adjustment of the GVD (and potentially even higher order phase terms) to maintain a short pulse duration is expected to increase the gain even further similar to the behaviour described before.\\
An appropriate adjustment of the GVD (and potentially even higher order phase terms) to maintain a short pulse duration is expected to increase the gain even more as demonstrated in FIG.~\ref{fig:gvd_rcf}.\\
The described spectral phase term optimization was subsequently established as a daily preparation routine during proton acceleration experiments at the DRACO laser.
%Data recorded during this optimization procedure over a period of more than one year of operation is presented in FIG.~\ref{performance}.
On the basis of a few shots each day, the best performing GVD-TOD value combination was evaluated and then applied for the rest of the experiment.
Optimal TOD values ranged between $20\,-40\mathrm{k}\,\mathrm{fs^3}$ and GVD values were adapted accordingly so that the pulse duration became minimal.
Maximum proton energy data recorded on 45 different shot-days (575 total shots) over a period of more than one year of operation is presented in FIG.~\ref{performance}.
For the standard settings of the spectral phase, it can be seen that the maximum proton energy for individual shots is fluctuating between 25\,MeV and 65\,MeV, resulting in an average energy of $(42.6\pm9.1)$\,MeV.
When changing to the optimized settings, maximum energies fluctuate between 40\,MeV and 71\,MeV, with reduced shot-to-shot fluctuation and an increased average energy of $(58.2\pm6.2)$\,MeV.
The red solid curve in FIG.~\ref{performance} shows the performance enhancement between the standard and optimized conditions, yielding an average cut-off energy gain of $\sim 37\,\%$.
%As shown by the red curve in FIG.~\ref{performance},  a performance enhancement is present in most cases, often yielding a gain of more than 50\% in cut-off energy.
%Also, the shot-to-shot fluctuation is reduced when the optimized setting is applied.
%
%

%\sout{Existing literature simulating asymmetric pulse shapes is contradictory \cite{Souri2017,Kumar2019} and shows much lower gain values.
%This was confirmed by particle-in-cell simulations that asymmetric pulse shapes for our specific experimental parameters, do by far underestimate the experimentally observed gain in particle number and energy.
%This was confirmed by particle-in-cell simulations for our specific experimental parameters and asymmetric pulse shapes. The results do by far underestimate the experimentally observed gain in particle number and energy.
%We further found that variations in the rising edge in the few hundreds fs range or post pulses also do not change the final proton energies by more than 10\% (see Supplementary for more details on the performed simulations).
%Based on those simulations we conclude that a more detailed understanding of the temporal plasma evolution coupled to the performed spectral phase changes of the laser is necessary.
%One major challenge that needs to be tackled in the future is the determination of the real plasma conditions several ps before the main pulse arrival, which means that the full energy laser pulse contrast and the corresponding plasma response need to be precisely known.}
%
%------------------------------------------
\section*{Conclusion}
%------------------------------------------
In conclusion, this paper shows that temporal pulse modification enables application relevant proton beam quality with a state-of-the-art high-repetition rate PW laser system.
Using an AOPDF and manually manipulating the spectral phase, notably the third order dispersion term, significantly enhances proton acceleration up to 70\,MeV.
%last picoseconds of the temporal intensity contrast and especially to the symmetry of the laser pulse shape.
The highest proton cut-off energies were measured for temporal pulse parameters well different from those of ideally compressed FTL pulses.
%Short, asymmetric laser pulses with a shallow rising edge can effectively double the maximum proton energy and significantly increase the particle flux as well.
Optimal pulse shape modification can lead to a significant gain of maximum proton energy and particle numbers.
The demonstrated stability of this effect during a long period of operation and over a wide range of parameters like target thickness and material as well as laser energy and temporal intensity contrast implies that this method could be easily transferred to 
other laser systems operating in the PW range. 
%Existing literature simulating asymmetric pulse shapes is contradictory \cite{Souri2017,Kumar2019} and shows lower gain values.
Existing literature simulating asymmetric pulse shapes \cite{Souri2017,Kumar2019} show gain values from 50\,\% to 65\,\%, but is not conclusive about the required type of asymmetry.
Moreover, they do not cover the experimental parameters of this study as they were performed in a different regime.
%Double pulse structures \cite{Brenner2014,Ferri2018} may explain the effect only for cases where the second pulse contains the majority of the laser pulse energy, which disagrees with the pulse shape induced by positive TOD values.
%Double pulse structures \cite{Brenner2014,Ferri2018} may explain the effect only for cases where the second pulse contains either the same amount or the majority of the laser pulse energy, which disagrees with the pulse shape induced by positive TOD values.
Double pulse structures \cite{Brenner2014,Ferri2018} may explain the effect only for cases where the second pulse contains either the same amount or the majority of the laser pulse energy, which disagrees with the pulse shape induced by positive TOD values.
Therefore, numerical investigations to unravel the complete microscopic picture of the complex laser-plasma acceleration process when using non-perfectly compressed pulses are necessary.
As an important result of the presented spectral phase manipulation technique, further experimental and numerical research should focus on all different aspects of the thereby induced laser intensity distribution changes within the complete pico-second time window around the main pulse to further improve control of the observed enhancement effect.
One major challenge to be addressed for this in the future is the determination of the real plasma conditions several ps before the main pulse arrival, which means that the full energy laser pulse contrast and the corresponding plasma response need to be precisely known.
% Dies erfordert ggf erst noch eine Weiterentwicklung der aktuellen Kapazitäten von Experiment und Simulation --> komplexität klar machen und warum wir das noch nicht lösen konnten
Note, in perspective of future applications, automated dispersion control to optimize laser proton acceleration is a readily applicable method to be combined with real-time feedback routines based on advanced computing schemes.\\
\newline
The data that support the findings of this study are available
from the corresponding author upon reasonable request.
\bibliography{tz_bib}
%---------------------------------------

%For data citations of datasets uploaded to e.g. \emph{figshare}, please use the \verb|howpublished| option in the bib entry to specify the platform and the link, as in the \verb|Hao:gidmaps:2014| example in the sample bibliography file.

\section*{Acknowledgements}

We gratefully acknowledge the DRACO laser team for excellent experiment support.
The work was partially supported by H2020 Laserlab Europe V (PRISES, Contract No. 871124), as well as the QST President's Strategic Grant (QST International Research Initiative (AAA98) and Creative  Research (ABACS)) and by the JST-MIRAI Program (Grant No. JPMJMI17A1, Japan).
M.N. was supported by JST PRESTO Grant No. JP-MJPR16P9 and by the Mitsubishi Foundation.
\section*{Author contributions statement}
T.Z., C.B., F.-E.B., S.K., F.K., J.M., L.O., M.R., H.-P.S. and K.Z. have set up and conducted the experiment. T.Z., S.B., R.G, U.H., A.I., T.P., U.S and K.Z. contributed to laser operation and diagnostics development.
T.Z. analyzed the data and prepared the figures.
T.Z and K.Z. wrote the manuscript.
T.C., U.S. and K.Z. supervised the project. All authors reviewed the manuscript and contributed to discussions.
%
%--------------------------------------------
\section*{Additional information}
%--------------------------------------------
%To include, in this order: \textbf{Accession codes} (where applicable);
\textbf{Competing interests:} The authors declare no competing interests.
\end{document}